\documentclass[11pt]{physmath_art}

\usepackage{cite} % Make references as [1-4], not [1,2,3,4]
\usepackage{url}  % Formatting web addresses  
\usepackage{ifthen}  % Conditional 
\usepackage{multicol}   %Columns
\usepackage[utf8]{inputenc} %unicode support
\usepackage[dvips]{graphicx}

\def\lim{\mathop{\rm lim}}

\newboolean{publ}

\newenvironment{physmath}{\begin{raggedright}\baselineskip16pt\setboolean{publ}{false}}{\end{raggedright}}

\begin{document}
\begin{physmath}

\title{Supernova Ia observations in the Lema\^itre--Tolman model}

\author{Krzysztof Bolejko\correspondingauthor$^{1}$%
       \email{Krzysztof Bolejko\correspondingauthor - bolejko@camk.edu.pl}%
      }

\address{%
    \iid(1)Nicolaus Copernicus Astronomical Center, Bartycka 18, 00-716 Warsaw, Poland
}%

\maketitle

 \begin{abstract}
  \parbox{14cm}
		{
The aim of this paper is to check if the models with realistic inhomogeneous matter distribution and without cosmological constant can explain the dimming of the supernovae in such a way that it can be interpreted as an acceleration of the Universe. 
Employing the simplest inhomogeneous model, i.e. Lema\^{i}tre--Tolman model, this paper examines the impact of inhomogeneous matter distribution on  light propagation. 
These analyses show that realistic matter fluctuations on small scales  induce brightness fluctuations in the residual Hubble diagram of amplitude around $0.15$ mag, and thus can mimic  acceleration. However, it is different on large scales. All these brightness fluctuations decrease with distance and hence cannot explain the dimmining of supernovae for high redshift without without invoking the cosmological constant.This paper  concludes that models with realistic matter distribution (i.e. where variation of the density contrast is similar to what is observed in the local Universe) 
cannot  explain the observed dimming of supernovae without the cosmological constant. \pb
		}

\end{abstract}

\section{Introduction}

This paper examines  the supernova observations in order to  thoroughly estimate the influence of inhomogeneities on light propagation. Studies in this field proved that inhomogeneities can mimic the cosmological constant. However, this does not prove consistent with other astronomical observations. This paper  provides some  quantitative insight
to matter fluctuations' influence in terms of the amplitude, $\delta m$,  measured in the residual Hubble diagram.

The observations of supernovae are a powerful tool in   modern cosmology. Analyses  of 
the supernova brightness provide us with a reliable estimation of their distance from an observer. For this estimation to be satisfactory, all factors which might influence the observed supernova luminosity must be taken into account.
In literature five factors are examined; namely, evolution of supernovae, dust absorption, selective bias, gravitational lensing, and cosmological models. Except for the last one they do not seem to be responsible for observed 'dimming' (for details see
Refs. \cite{Fil, Ris, Tor, Pr1, Pr2}). Analyses of supernovae in various homogeneous cosmological models imply a non-zero cosmological constant. 
However, similar analyses in inhomogeneous models have not been systematically studied.

The luminosity distance of supernovae without the homogeneity assumption is to be analysed as well.
The luminosity distance in inhomogeneous models might differ from the FLRW results.
To examine this issue  the Lema\^{i}tre--Tolman model is employed.
In this approach not only matter is inhomogeneously distributed but  the expansion of the space in not uniform as well.
Results in the form of  the residual Hubble diagram provide us with the estimation of the impact of matter inhomogeneities.

The effect of inhomogeneous matter distribution on supernova observations
was studied by many authors.
For example employing the Lema\^itre--Tolman model and the Taylor expansion of the luminosity distance in powers of the redshift C\'el\'erier \cite{MNC1} showed that the inhomogeneities
can mimic the cosmological constant.
Iguchi, Nakamura and Nakao \cite{INN} also used 
the Lema\^itre--Tolman model to show that it is possible to fit supernova data without
the cosmological constant. Similar results were obtained 
in the Stephani model by God\l{}owski, Stelmach and Szyd\l{}owski \cite{GSS}.
and in the Szafron model \cite{Moffat}. Also models by 
Alnes, Amarzguioui and Gron \cite{AAG} where the density is increasing with distance
and models by Enqvist and Mattsson where expansion is decreasing with distance \cite{EM}
successfully fit supernova data without a need for the cosmological constant.
There have also been other models proposed,
in particular Swiss cheese models by Mansouri \cite{Man} and 
Brouzakis, Tetradis and Tzavara \cite{BTT1,BTT2}. For a review 
on explanation of the acceleration expansion without the cosmological constant the reader is referred to Ref. \cite{MNC2}.
The effect of inhomogeneities was also studied with aid of approximate methods \cite{GCA,KMNR,KMR}.
Recently Vanderveld, Flanagan and Wasserman \cite{VFW} 
studied this issue using perturbation approach up to the second order in density fluctuations.
Their results are similar to \cite{GCA,KMNR,KMR} and indicate that
the effect of inhomogeneity on the expansion of the Universe  
is small and thus cannot explain the apparent acceleration.
However, because of the perturbation framework
their results are valid only for small values of density fluctuations.
Since the real density fluctuations in our Universe largely exceed $\delta \sim 1$
in order to draw reliable conclusion similar analyses should be conducted 
by employing exact solution of the Einstein equations.

These studies  have shown that matter inhomogeneities can explain the apparent acceleration of our Universe without employing the cosmological constant. This paper not only indicates that there are some specific conditions which enable explanation of the supernova dimming without $\Lambda$ but also examines the influences of the realistic matter distribution on light propagation.

The structure of this paper is as follows:  Sec. \ref{ltt} 
presents the Lema\^itre--Tolman model; in  Sec. \ref{obscon} presents observational constraints; Sec. \ref{wlrfl} presents the residual Hubble diagram for models with  realistic density distribution but without the cosmological constant; Sec. \ref{fita} presents results of fitting models to the supernova measurements without the cosmological constant; Sec. \ref{ccl} presents the residual Hubble diagram for models with the realistic density distribution and with the cosmological constant.

\section{The Lema\^{\i}tre-Tolman  model}\label{ltt}
   
 The Lema\^{\i}tre-Tolman \cite{Lem,Tol}
 model is a spherically symmetric solution of the Einstein   
equations with a dust source.
In comoving and synchronous coordinates,  the metric is:   
\begin{equation}   
{\rm d}s^2 =  c^2{\rm d}t^2 - \frac{R'^2(r,t)}{1 + 2 E(r)}\ {\rm   
d}r^2 - R^2(t,r) {\rm d} \Omega^2, \label{ds2}   
\end{equation}   
where $ {\rm d} \Omega^2 = {\rm d}\theta^2 + \sin^2 \theta {\rm d}\phi^2$,    
and $E(r)$ is an arbitrary function of $r$. Because of the signature   
$(+, -, -, -)$, this function must obey $E(r) \ge - \frac{1}{2}.$   
   
The Einstein  equations can be reduced  and presented as follows:   
\begin{equation}\label{den}   
\kappa \rho c^2 = \frac{2M'}{R^2 R'},   
\end{equation}   
\begin{equation}\label{vel}   
\frac{1}{c^2} \dot{R}^2 = 2E(r) + \frac{2M(r)}{R} + \frac{1}{3} \Lambda R^2,   
\end{equation}   
\noindent where $M(r)$ is another arbitrary function and $\kappa = \frac{8 \pi G}{c^4}$.   
   
When $R' = 0$ and $M' \ne 0$, the density becomes infinite. This    
happens at shell crossings. 
This is an additional singularity to the Big Bang that occurs at $R = 0, M' \neq 0$.
The shell crossing can be avoided by setting the initial conditions appropriately \cite{HL}.

Equation (\ref{vel}) can be solved by a simple integration:   
    
\begin{equation}\label{evo}   
\int\limits_0^{R(r)}\frac{ {\rm d} \tilde{R}}{\sqrt{2E(r) + \frac{2M(r)}{\tilde{R}} + \frac{1}{3}\Lambda \tilde{R}^2}} = c    
\left(t- t_B(r)\right),    
\end{equation}   
where $t_B(r)$ appears as an integration constant, and is an arbitrary   
function of $r$. This means that unlike in the Friedmann models the Big Bang is not a single event,
but it can occur at different times at different distances from the origin.
   
Thus, the evolution of the Lema\^{\i}tre--Tolman model is determined by three arbitrary   
functions: $E(r)$, $M(r)$, and $t_B(r)$. The metric and all the formulae   
are covariant under arbitrary coordinate transformations of the form $r   
= f(r')$. Using such a transformation, one of the functions determining the Lema\^{\i}tre-Tolman model can be given a   
desired form. Therefore, the physical initial data of the  Lema\^{\i}tre-Tolman   
model evolution consists of two arbitrary functions (see Ref. \cite{HK2006} on how to specify the Lemia\^{i}tre--Tolman model).

\subsection{The Hubble parameter}

In the Friedmann limit $R \rightarrow r a$ [$a(t)$ is the scale factor], so the simplest generalisation of the Hubble constant which in Friedmann models is $H_0 = \dot{a}/a$ would be $H=\dot{R}/R$.
However, from the comparison of the  approximate distance-redshift relation  
\cite{Ell,PK} with the Hubble law, the Hubble parameter
would rather be $H=\dot{R}'/R'$. However, the above mentioned relation is valid only for low
redshift and thus if one analyses astronomical data of high redshift one should refer
to the definition of the Hubble constant  based on the rate of volume change \cite{Ell},
i.e. $H= (1/3) \Theta$ (where $\Theta$ is the scalar of the expansion), which 
in the Lema\^{i}tre--Tolman model is:

\begin{equation}
H = \frac{1}{3} \Theta = \frac{1}{3} \left( 2 \frac{\dot{R}}{R} + \frac{\dot{R}'}{R'} \right). 
\label{hldf}
\end{equation}

The above  ambiguity in the definitions of the Hubble constant shows 
that one should be very careful in measuring the parameter which is called
the Hubble constant.  For example the value of the Hubble constant derived
from the continuity equation:
$H_c =  -(1/3) (\dot{\rho} / \rho$) (where $\rho$ is density) is in general not equal to the 
Hubble constant derived from the measurements of the Hubble flow:
$H_f = V/D$ (where V is the receding velocity, D distance).

In this paper when referring to the Hubble parameter it is assumed that
the Hubble parameter is defined by eq. (\ref{hldf}).
Please note that at origin because of the regularity condition
\cite{KH} $R \rightarrow ra$ and $H$ coincides with $H_0$.

\subsection{The redshift formula}

Light propagates along null geodesics. The vector tangent
to the null geodesic, $k^{\alpha}$, obeyes the following relation:

\begin{equation}
k^{\alpha}{}_{; \beta} k^{\beta} = 0.
\label{geoeq}
\end{equation}

As light propagates the frequency of photon changes. The ratio
of the frequency at the emission instant to the frequency at the observation moment
defines the redshift:

\begin{equation}
\frac{\nu_e}{\nu_o} := 1+z.
\end{equation}

The energy of  photons measured by an observer with a 4--velocity 
$u^{\alpha}$ is proportional to $k^{\alpha} u_{\alpha}$.
Thus, the redshift formula is as follows:

\begin{equation}
1 + z = \frac{\left( k^{\alpha} u_{\alpha} \right)_e}{
\left( k^{\alpha} u_{\alpha} \right)_o}
\label{redlaw}
\end{equation}
where subscripts $_e$ and $_o$ refers to instants
of emission and observations respectively.

In the Lema\^itre--Tolman model the above formula reduces to \cite{Bon,PK}:

\begin{equation}
\ln (1 +z) = \frac{1}{c} \int\limits_{r_{o}}^{r_{e}} {\rm d} r \frac{\dot{R}'(t(r),r)}{\sqrt{1+2E(r)}},
\label{redlt}
\end{equation}
where all the above quantities  are evaluated at the null cone,
i.e. they can be calculated by solving the following equation:

\begin{equation}
c {\rm d} t = - \frac{R'(t,r)}{\sqrt{1+ 2E(r)}} {\rm d} r.
\label{geo}
\end{equation}

\noindent From eq. (\ref{vel}) we obtain:

\begin{equation}
\dot{R}' = \frac{c^2}{\dot{R}} \left( \frac{M'}{R} - \frac{M R'}{R^2} + E' + \frac{1}{3} \Lambda R R' \right).
\label{rrt}
\end{equation}

Models considered in this paper were defined by functions presented in Table \ref{Tab1}.
The radial coordinate was chosen as a present day value of the areal radius, i.e.
$r :=  R_0$. 
In the case when model is defined by a pair $t_B$ and $\rho$, $M(r)$ is calculated from 
eq. (\ref{den}) then the function $E(r)$ is calculated from eq. (\ref{evo}).
In the case when model is defined by a pair $H(r)$ and $\rho$, $M(r)$ is calculated from 
eq. (\ref{den}), then $\dot{R}$ is calculated from eq. (\ref{hldf}), and finally 
eq. (\ref{vel}) is used to calculate  $E(r)$.

Once the functions $M(r), E(r)$ are known, the evolution equation 
 --- eq. \ref{vel} --- can be solved and the evolution of the model can be traced back in time.
 Simultaneously eq. (\ref{geo}) is solved in order to calculate all quantities at the null cone. Then, using eqs. (\ref{rrt}) and (\ref{redlt}) the redshift can be estimated.
Finally, from the  reciprocity 	theorem \cite{Ell}, the luminosity distance is calculated using the following relation:

\begin{equation}
D_L(t(r),r) = R(t(r),r) (1+z)^2.
\end{equation}

\section{Observational constrains}\label{obscon}

The astronomical observations providing us with  information about the local Universe
prove that  matter distribution and  expansion of the space are not homogeneous. 

The measurement of the matter distribution  implies that the density contrast ($\delta = \rho/\rho_b - 1$) varies from $\delta \approx -1$ in voids \cite{Hoy} to  $\delta$ equal to several tens in clusters \cite{Bar}. These structures are of diameters varying  from several Mpc up to several tens of Mpc.
  However, if averaging is considered on large scales, the density varies from 
  $0.3 \rho_b$ to $4.4 \rho_b$ \cite{Kol,Hud} and the structure sizes  are of several tens of Mpc.  
So far there is no observational evidence that structures larger than 
supercluster, i.e. of diameters of hundreds of Mpc or larger exist in the Universe.  

The measurements of the Hubble constant provide us with diffrent values of $H_0$ --- from $61 \pm 3 $~(random) $\pm~ 18$ (systematic)  km s$^{-1}$ Mpc$^{-1}$ \cite{Res}, to $H_0 = 77 \pm 4 \pm 7 $ km s$^{-1}$ Mpc$^{-1}$ \cite{Try}.  
 However, due to very large observational and systematical errors (larger than $10\%$) it is impossible to observe any  variations of the Hubble constant.

 This paper assumes that any realistic model must remain consistent with the above astronomical data. Namely, in models with the Hubble parameter as a variable we expect these variations to stay within the range indicated by the above observations. 
 Analogously, in models with an inhomogeneous density distribution, we expect the density fluctuations to remain within the range indicated by the observations.

\section{Results}\label{resul}

The supernova observations are provided by the gold data set \cite{Ris}.
This data is presented in form of the
distance moduli, i.e.:

\begin{equation}
\mu = m - M = 5 \log D + 25
\label{dm}
\end{equation}
where $m$ represents an observed magnitude, $M$ --- absolute magnitude, and $D$ ---  luminosity distance expressed in Mpc. The usual way of presenting the supernova data is the residual Hubble diagram. The residual Hubble diagram presents $\Delta m$ as a function of redshift:

 \begin{equation}
\Delta m = m - m^{emp} = 5 \log \frac{D}{D^{emp}}.
\label{rhd}
\end{equation}
where $m^{emp}$ is an expected magnitude in an empty RW model.
 
The luminosity distance in the empty cosmology is larger than in the decelerating FLRW universe but it is smaller than in the accelerating FLRW universe.
Therefore, if the supernovae are fainter (of higher magnitude) than they would be for an empty universe, this is interpreted as an evidence of acceleration.
In the analyses below the results are presented in the form of the residual Hubble diagrams.
The chosen background model, on which fluctuations will be imposed, is the FLRW model with the density:

 \begin{equation}
 \rho_b = 0.27 \times \rho_{cr} = 0.27 \frac{3H_0^2}{8 \pi G},
  \label{rbdf}
  \end{equation}
and the Hubble constant $H_0 = 65$ km s$^{-1}$ Mpc$^{-1}$.

\subsection{Models without cosmological constant}\label{without}

This section examines if the observed dimming of the supernova brightness 
may be caused merely by the matter inhomogeneities, without employing the cosmological constant. 
To do so the cosmological constant is set to zero and presureless matter is assumed to be the only component of the Universe.

\subsubsection{Realistic fluctuations}\label{wlrfl}

Astronomical observations of the local Universe indicate that its density varies from low values in voids to high values in clusters. 
Models 1, 2 and 3 are rough estimates of this phenomenon.
In model 1 the majority of regions through which supernova light propagates are of low density. In model 2 most regions' density is higher than the background density.
Model 3 has a cosine variation of  density  and its average density
 is of the background value.
The exact form of these fluctuations is presented in Table \ref{Tab1}.
Alhtough the density distribution in above models is spherically symmetric
and the real matter distribution in the Universe is not, such estimation
is adequate if the time of the light propagation is small.
For larger periods of time the evolution of  matter becomes important.
However since redshift $z \approx 0.5$ the Universe did not evolve
significantly, so up to redshifts $z \approx 0.5$ the analysis
presented here should not differ significantly form 
reality. For higher redshifts we may expect larger differences between
the results of these models and the real picture.
Despite these differences, such analysis is important
because it provides us with estimation
of the influence of light propagation effects  on the final results
of supernova observations.

Current density distributions are equal to these shown in Fig. \ref{dend123}.
Note that this graph only represents density up to 200 Mpc
 to demonstrate a periodic character of assumed density distributions.
As  mentioned in Sec. \ref{ltt} to specify the Lema\^itre--Tolman model two initial conditions have to be known.
The first initial condition is the density distribution.
The second initial data in this section is the distribution of the {\it bang time} function. It is assumed that $t_B(r) = 0$.
This assumption follows from the Cosmic Microwave Background (CMB) observations. These observations imply that the Universe was very homogeneous at the last scattering moment and as a consequence the amplitude of the {\it bang time} function could not be larger than a few thousand  years, which in comparison with the present age 
of the Universe is negligible.
If the $t_B$ were of larger value 
the temperature fluctuations would be greater than it is observed in the CMB sky \cite{BKH}.

Using the algorithm from Sec. \ref{ltt} the luminosity distance was calculated for the three above mentioned models. 

The results are presented in  the form of the residual Hubble diagram in Fig. \ref{resh123}   and indicate
 that realistic density fluctuations can mimic  the acceleration  on small scales. 
  Firstly, in the residual diagram there are some regions where $\Delta m$ is positive. Secondly,
  in some regions
the luminosity distance increases faster  than in the FLRW models. However, on large scales, a tendency for curves to decrease remains unchanged.
Near the origin the fluctuations in residual diagram are large
and are approximately equal to $0.15$ mag, but they are decreasing with  distance. 

Fig. \ref{resh123} also depicts the curve for the homogeneous $\Omega_{mat} = 0.27, ~\Omega_{\Lambda} = 0$ model
which, however, is not clearly visible due to very tight fluctuations of model 3 around it.
Curve 2 presented in Fig. \ref{resh123} lies  above the curve of the the homogeneous hyperbolic model because in this model the expansion of the space is smaller than the expansion of the homogeneous Universe. This is because in this model the density of regions through  which light propagates is larger than the background density. In model 1, a vast majority of the region is of lower density, hence, curve 1 is below the curve representing the hyperbolic homogeneous Universe.
As one can see, if $\Lambda = 0$, the realistic density fluctuations alone cannot be responsible for the observed dimming of the supernova brightness.

\begin{figure}
\begin{center}
\includegraphics[scale=1]{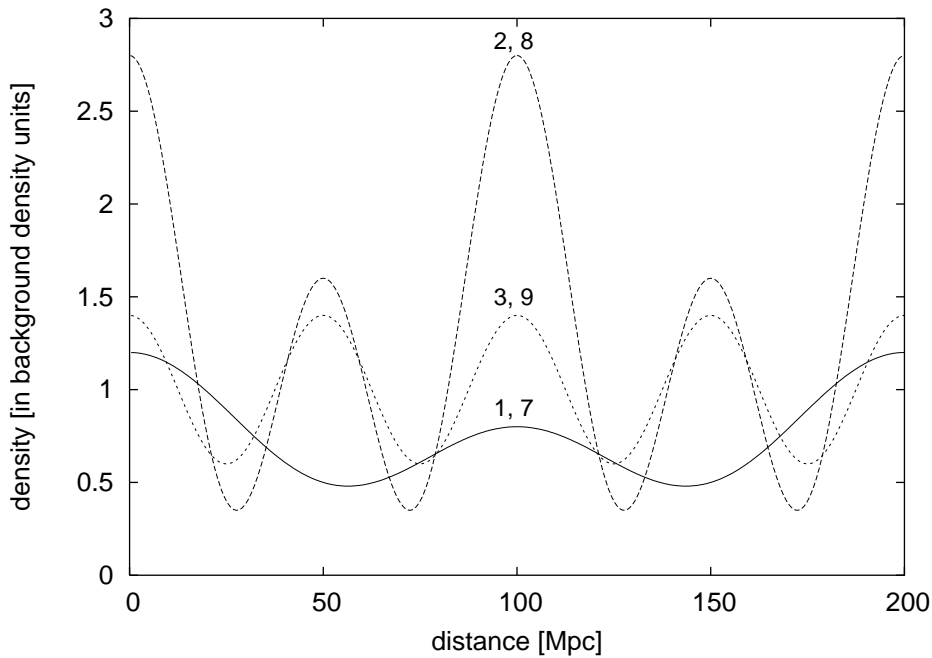}
\caption{Density distribution models 1, 2 and 3 (Sec. \ref{wlrfl}), and models 7, 8, 9 
(Sec \ref{ccl}).}
\label{dend123}
\end{center}
\end{figure}  

\begin{figure}
\begin{center}
\includegraphics[scale=1]{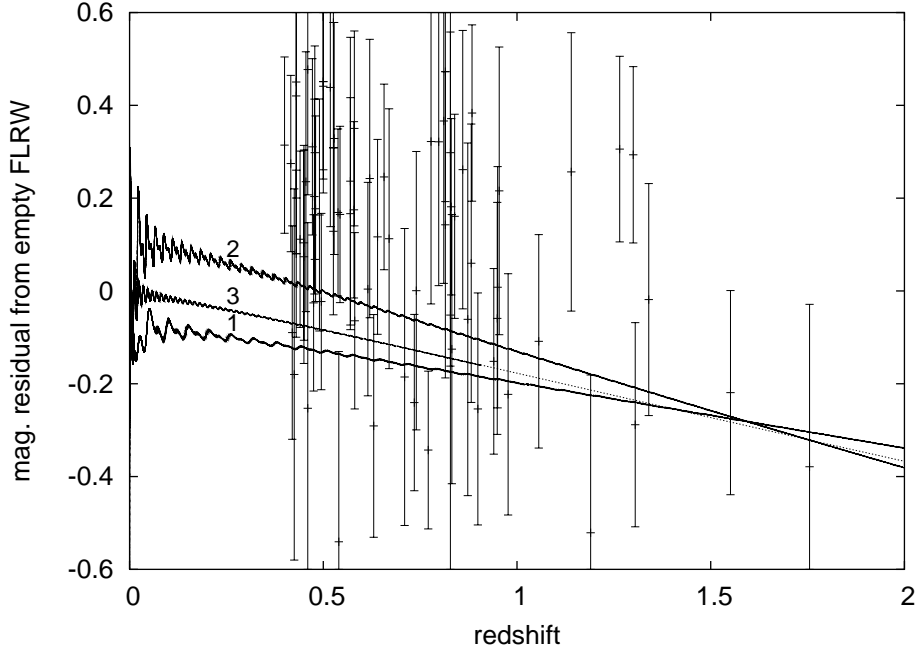} 
\caption{Magnitude residual diagram for models 1, 2 and 3. For clarity  supernova data (gold data set) from Ref. \cite{Ris} is only presented  for redshifts larger than 0.4.}
\label{resh123}
\end{center}
\end{figure} 

\begin{table}
\begin{center}
 \caption{The exact form of functions used to define models 1--9.}
\label{Tab1}
  \begin{tabular}{lc}
 Model  & Pair of functions \\ \hline
  model 1 &  $t_B = 0$; $\rho/\rho_b = 0.5  + 0.2\cos (10^{-5} \pi r  {\rm Mpc}^{-1}) $ \\ 
& $+ 0.5\cos^2 (10^{-5} \pi r {\rm Mpc}^{-1}) $ \\
 model 2 &   $t_B = 0$; $\rho/\rho_b = 0.4 + 0.6 \cos (2 \times 10^{-5}  \pi r {\rm Mpc}^{-1})  $\\
& $ + 1.8\cos^2 (2 \times 10^{-5} \pi r  {\rm Mpc}^{-1} )$ \\
 model 3 &  $t_B = 0$; $\rho/\rho_b = 1 + 0.4\cos (10^{-5} \pi r   {\rm Mpc}^{-1})$ \\
 model 4 &  $t_B = 0$; $\rho/\rho_b = 1 + (8 \times 10^{-6} r {\rm Mpc}^{-1})^{0.55}$ \\
model 5 &  $\rho/\rho_b = 1$; \\
&  $H/H_0$ --- not an analytic functions, see Fig. \ref{hubp45}  \\
model 6 &  $t_B = 0 $; $H/H_0  = 1$ \\
model 7 &  as in model 1 \\
model 8 &  as in model 2 \\
model 9 &  as in model 3 \\
\end{tabular}
\end{center}
\end{table}

\subsubsection{Fitting the observations}\label{fita}

It has been proved that the Lema\^itre--Tolman model may be fitted to any set of observational data \cite{MHE}. Thus, the Lema\^itre--Tolman model can always be fitted to supernova data, without employing a cosmological constant.
Nevertheless, if such a fitted model is in consistent with all the astronomical data (such as galaxy redshift surveys, CMB), then the problem remains unresolved. 
This section addresses the above mentioned problems.

To specify a Lema\^itre--Tolman  model one needs to know two initial functions.
The functions such as $E(r)$ or $M(r)$ are difficult to extract from observations. However, the observations provide us with the measurements of $\rho$, $H_0$ and $t_B$.
In this section these functions are chosen to enable one of them to be consistent with the astronomical observations,  while the second function remains in accordance with the supernova observations as much as possible.

The Following models are considered:
\begin{enumerate}
\item Model 4.
 
 In  model 4 the bang time function $t_B(r) = 0$ is consistent with the CMB observations. The density distribution is chosen  so that it fits the supernova  observations. The results are presented in Fig. \ref{resh45}.  The values of $\chi^2$ test are presented in Table \ref{Tab2}.
  The density distribution in model 4  monotonically increases;  from an average value ($\rho = \rho_b$) at the origin to a value of $\rho = 2.5 \rho_b$ at the distance of $3$  Gpc.  The increase of density yields a decrease of the expansion.
Fig. \ref{hubp45}. presents the Hubble parameter (as defined by eq. (\ref{hldf})).

 If local density  and Hubble flow measurements extend up to the distance of Gpc then
  it might be supposed that 
  model 4 is  unrealistic. However, there are no systematic observations of the density distribution or expansion at distances of Gpc and all that is really known is that the 
     relative motion of our Galaxy with respect to the CMB is small. This implies that to explain the relatively small motion with respect to the CMB rest frame  the expansion of the Universe should increase at a larger distace. 
As can be seen the Lema\^itre--Tolman model is of a great flexibility so one can always choose such functions which would  fit the CMB (the diameter distance to
the surface of the last scattering is approximately 14 Gpc). This, however,  requires futher complications of such a model.

\item Model 5.
 
 In  model 5 the density distribution is assumed to be equal to the background value $\rho = \rho_b$. This implies that no Gpc--scale structures exist in it. The second function which defines the Lema\^itre--Tolman model
is the Hubble parameter which is  chosen  to fit the supernova observations. The variations of the Hubble parameter are presented in Fig. \ref{hubp45}. 
   
  As can be seen in Fig. \ref{hubp45}  the variations of the Hubble parameter are comparable within $3 \sigma$ estimation of the Hubble constant. However,  such a behaviour is rather unrealistic, and  together with
   model 4  
  suggests the existence of very large scale structures (of several Gpc diameters). 
       Furthermore, in model 5 the {\it bang time} function is very inhomogeneous and decreases to almost $-1.7$ billions years at distance of $2.4$. 
   Such a large amplitude of $t_B$ is strongly inconsistent with CMB observations.

\item Model 6

In this model the Hubble parameter is chosen to be of $65$ km s$^{-1}$ Mpc$^{-1}$. The density and $t_B(r)$ are chosen to fit the supernova data.
However, none of the attempts to obtain a satisfactory fit to the observational data succeeded.
The best fitted model within the family of constant H models  is the empty FLRW model (with $\Delta m = 0$). 

\end{enumerate}

\begin{figure}
\begin{center}
\includegraphics[scale=1]{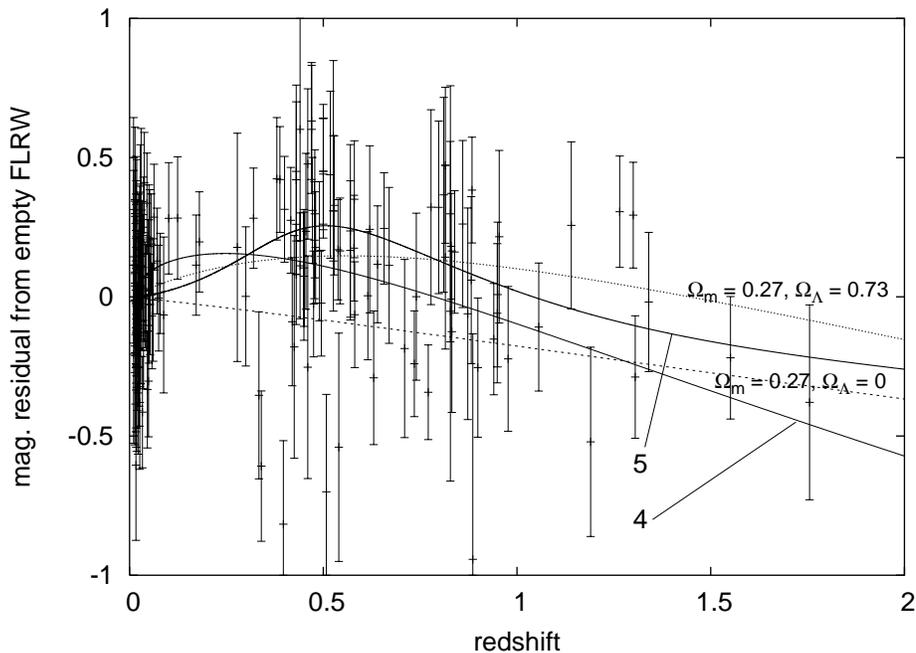}
\caption{The residual Hubble diagram for models 4 and 5.}
\label{resh45}
\end{center}
\end{figure} 

\begin{figure}
\begin{center}
\includegraphics[scale=1]{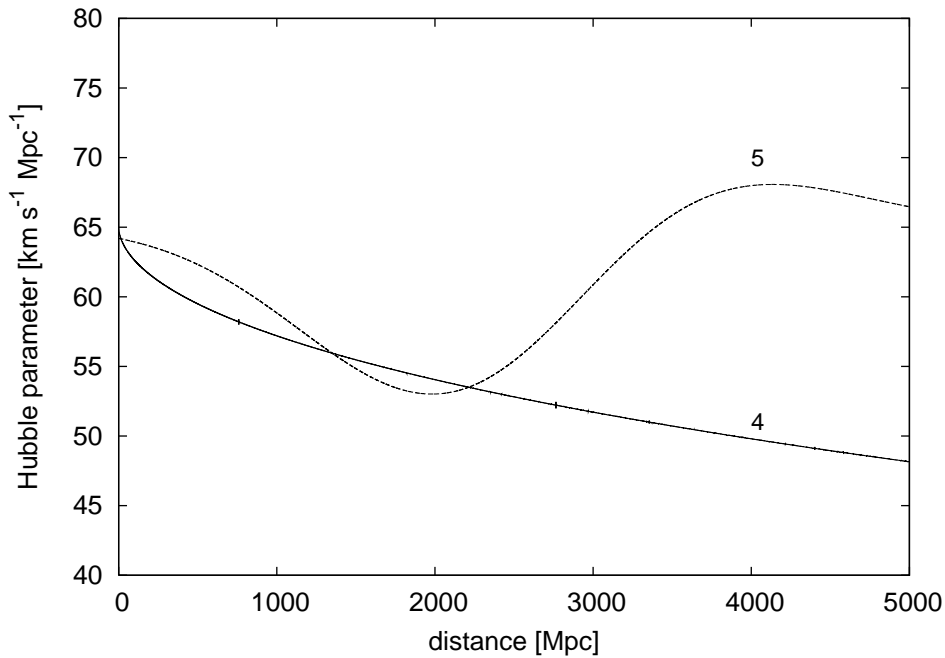} 
\caption{The Hubble parameter for models 4 and 5.}
\label{hubp45}
\end{center}
\end{figure}

The above results suggest
that the only way to fit the supernova data
is to set the expansion of the Universe to be decreasing on the past null--cone.
This can be done either be setting the expansion of the Universe to be decreasing with
radial coordinate (models 4 and 5) or to assume the existence
of cosmological constant (standard approach).
The first alternative implies that 
the cosmological constant is not needed but
our position in the Universe is  very special and 
that on the scales of Gpc there exist  large structure in the Universe.
The second alternative is that the models 
presented in this section support the present--day acceleration of the Universe as an explanation of the supernova observations. 
Within the type of models considered above it is impossible to fit the
supernova data with realistic matter distribution (i.e. where variations of the density contrast are similar to what is observed in the local Universe).
Currently, in terms of analyses of observations it seems that these two interpretations are equally probable. The difference is in the philosophical assumptions. The first interpretation requires that  our  position  in the Universe is a special one.
 This, however, cannot be proved right or wrong by any current  observations.
The galaxy redshift surveys, like SDSS or 2dFGRS, measure galaxis up to redshift only $z \approx 0.4$. On the other hand the CMB observation provide us with information
about the state of the Universe which is currently about 14 Gpc remote from us.
Because of flexibility of the Lema\^itre--Tolman model, models 4 and 5
can be fitted to the CMB data, simply by assuming that this Gpc--structure is compensated by outer regions and than the Universe is homogeneous.
This implies the existance of very a large structure in the Universe,
of diameters of order of Gpc.
The second interpretation is based on the assumption that our position in the Universe is not special at all and on large scale the Universe is homogeneous.
As mentioned above this  assumption cannot be verified by observation, however there are some theoretical results that support this statement. These are the 
Ehler-Geren-Sachs (EGS) \cite{EGS} theorem and `almost EGS
theorem' \cite{SME} which states that if 
anisotropies in the cosmic microwave background radiation  
are small for all observers then our Universe must be 'almost FLRW' on large scales.
Therefore, it seems that the interpretation that the cosmological constant
is of a non--zero value seems to be more probable.

\begin{table}
\begin{center}
 \caption{Test $\chi^2$ of fitting the supernova observations.}
\label{Tab2}
  \begin{tabular}{ll}
  Model  & $\chi^2_{NDF}$ \\ \hline
 model 1 &  2.05 \\
 model 2 &  1.46 \\
 model 3 & 1.62 \\
 model 4 &  1.19 \\
 model 5 & 1.15  \\
 model 7 & 1.35 \\
 model 8 & 1.26 \\
 model 9 & 1.14 \\
 FLRW ($\Omega_m = 0,~\Omega_{\Lambda} = 0 $)  &  1.35 \\
 FLRW ($\Omega_m = 0.27,~\Omega_{\Lambda} = 0.73) $  &  1.14 \\
 FLRW ($\Omega_m = 0.27,~\Omega_{\Lambda} = 0 $)  &  1.59 \\
\end{tabular}
\end{center}
\end{table}

\subsection{Cosmological constant}\label{ccl}

This section investigates  the 
 light propagation in the inhomogeneous universe with the cosmological constant. The value of the cosmological constant corresponds to the concordance value, $\Omega_{\Lambda} = 0.73$.
Investigated models include
model 7, 8, and 9.
These models' density distribution is similar to the density distribution of models 1, 2, and 3 respectively. 
The results of fitting these models to supernova data are presented in Table \ref{Tab2}.

Results presented in Fig. \ref{resh789} indicate that realistic matter fluctuations (as in the case with no  ${\Lambda}$) introduce fluctuations to the residual Hubble diagram. These fluctuations are large for low redshifts but decrease fast for high redshifts.
It can be seen from Table \ref{Tab2} that all models with $\Lambda$ fit  the supernova data better.
The residual Hubble diagram presented in Fig. \ref{resh789} shows that the influence of the density fluctuations is significant only for small redshifts. 
It is uncertain  whether this phenomenon is real or is just a consequence of the spherical symmetry assumption.
Within a small distance from the origin, spherical symmetry is valid but as the distance increases it becomes less accurate. 
Fig. \ref{resh789} indicates that the amplitude of the fluctuations in the residual Hubble diagram is decreasing with  redshift. 
This can be due to the evolution --- in the past, the density fluctuations were of a smaller amplitude, hence the lower amplitude of fluctuations in the residual diagram. However, the Universe has not evolved significantly since the redshit $z \approx 0.5$. so it might be possible that in non-symmetrical models the amplitude of the magnitude fluctuations would not decrease so fast as in our case.
To confirm this hypothesis the above calculations should also be repeated in the inhomogeneous nonsymmetrical model.
If it is confirmed it  could partly explain the large scatter of the supernova data
 which is currently believed to be
 caused by numerous factors, like observational errors or non--uniform absolute brightness of the supernovae.

\begin{figure}
\begin{center}
\includegraphics[scale=1]{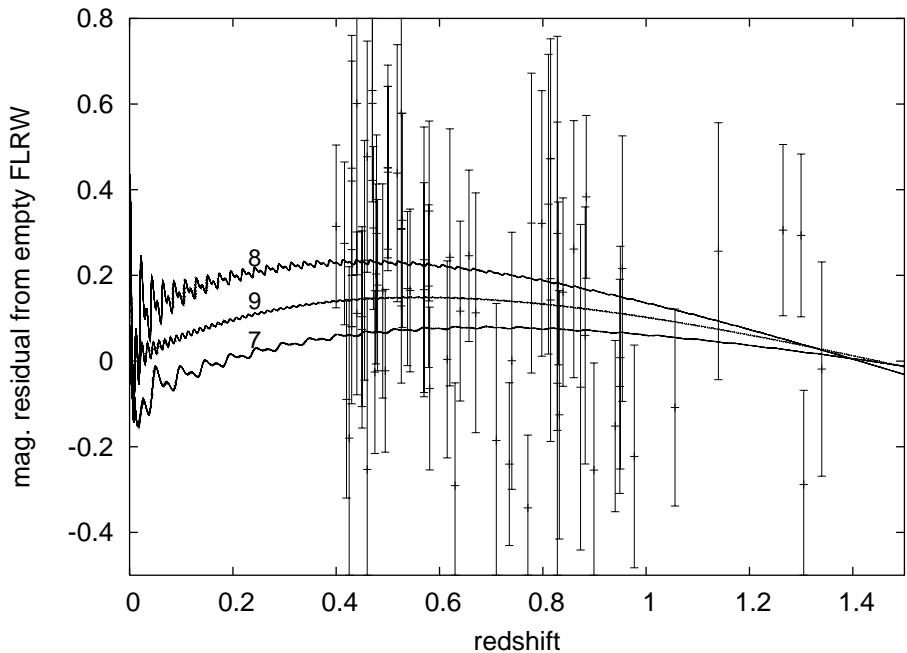}
\caption{Magnitude residual diagram for models 7, 8 and 9. For clarity reason  supernovae data (gold data set) from Ref. \cite{Ris} are presented only for redshift larger than 0.4.}
\label{resh789}
\end{center}
\end{figure}

\section{Conclusion}

This paper investigates 
the propagation of the light of supernovae  in the inhomogeneous Lema\^itre--Tolman model. The inhomogeneous models are of  great flexibility and can fit the data without invoking the cosmological constant, which has been proved by 
Mustapha, Hellaby and Ellis \cite{MHE}. Many authors before, like  
C\'el\'erier \cite{MNC1} or recently Alnes, Amarzguioui and Gron \cite{AAG} have
 proved that the matter inhomogeneities in the Lema\^itre--Tolman model can mimic the cosmological constant and thus can be an alternative to dark energy.
However, this paper indicates that
 the models which fit the supernova measurement without invoking the cosmological constant are very peculiar (see model 4 and 5, Sec. \ref{fita}). These models have either a very peculiar expansion of the space (decreasing from the origin), or an unrealistic density distribution (increasing from the origin) or/and a  very large amplitude of the bang time function ($t_B(r)$). Introducing the {\it Ockham's Razor} principle, it is more likely that the  Universe is accelerating rather than the conditions in our position in the Universe are so very special and
  extraordinary 
  that they could be possibly responsible for the observed dimmining of the supernova brightness.
 
 The results show that in the inhomogeneous Lema\^itre--Tolman model the amplitude of brightness fluctuations observed in the residual Hubble diagram is significantly large for low redshifts of amplitude around  0.15 mag but it decreases for higher redshifts. Thus, for redshifts larger than $z \approx 0.3$ these fluctuations are neglgible.
 All this may be the result 
  of the evolution (as
  in the past the density fluctuations were smaller, and, consequently were of smaller influence on the brightness fluctuations). However, it is also possible that this fast decrease can be due to the symmetry restrictions. The Lema\^itre--Tolman model assumes a spherical symmetry which puts too many constrains on the evolution and another parameters of the model. Therefore, it would be worth 
 investigating the light propagation in the models which are both non-symmetrical and inhomogeneous.
If in the inhomogeneous and non-symmetrical models the magnitude fluctuations do not decrease so fast, the observed scatter of supernova measurements might be partially possible to  explain. 

The main conclusion of this paper is that matter inhomogeneities introduce 
the brightness fluctuations to the residual Hubble diagram of amplitude approximately  $0.15$ mag for low redshifts, and thus can mimic the acceleration on small scales.
However, to explain the excess of faint supernovae without applying any special conditions (such as for instance peculiar expansion of the Universe) 
  the cosmological constant has to be employed.

\section*{Acknowledgements}
I am very grateful to  Charles Hellaby for cooperation, helpful discusions, and hospitality. I  would also like to thank Andrzej Krasi\'nski, Bill Stoeger and Paulina Wojciechowska for their   precious comments and help.  I thank the Department of Mathematics of the Cape Town University, where most of this research was carried out and NRF of South Africa for financing my visit.
This research has been partly supported by Polish Ministry of Science and Higher Education under grant N203 018 31/2873, allocated for the period 2006-2009.

\end{physmath}
\end{document}